# A topological Hund nodal line antiferromagnet

Xian P. Yang[1*], Yueh-Ting Yao[2], Pengyu Zheng[3], Shuyue Guan[4], Huibin Zhou[4], Tyler A. Cochran[1], Che-Min Lin[2], Jia-Xin Yin[5], Xiaoting Zhou[6,7], Zi-Jia Cheng[1], Zhaohu Li[8], Tong Shi[8], Md Shafayat Hossain[1], Shengwei Chi[8], Ilya Belopolski[1], Yu-Xiao Jiang[1], Maksim Litskevich[1], Gang Xu[8, 9, 10], Zhaoming Tian[8], Arun Bansil[6,7], Zhiping Yin[3,11], Shuang Jia[4,12,13], Tay-Rong Chang[2, 14, 15*], M. Zahid Hasan[1, 16, 17*]

[1]Laboratory for Topological Quantum Matter and Advanced Spectroscopy (B7), Department of Physics, Princeton University, Princeton, New Jersey, USA.
[2]Department of Physics, National Cheng Kung University, Tainan, Taiwan.
[3]School of Physics & Astronomy and Center for Advanced Quantum Studies, Beijing Normal University, Beijing, China.
[4]International Center for Quantum Materials, School of Physics, Peking University, Beijing, China.
[5]Department of Physics, Southern University of Science and Technology, Shenzhen, Guangdong, China.
[6]Department of Physics, Northeastern University, Boston, Massachusetts, USA.
[7]Quantum Materials and Sensing Institute, Northeastern University, Burlington, MA, USA.
[8]School of Physics and Wuhan National High Magnetic Field Center, Huazhong University of Science and Technology, Wuhan, Hubei, China.
[9]Institute for Quantum Science and Engineering, Huazhong University of Science and Technology, Wuhan, Hubei, China.
[10]Wuhan Institute of Quantum Technology, Wuhan, Hubei, China.
[11]Key Laboratory of Multiscale Spin Physics (Ministry of Education), Beijing Normal University, Beijing, China.
[12]Interdisciplinary Institute of Light-Element Quantum Materials and Research Center for Light-Element Advanced Materials, Peking University, Beijing, China.
[13]CAS Center for Excellence in Topological Quantum Computation, University of Chinese Academy of Sciences, Beijing, China.
[14]Center for Quantum Frontiers of Research and Technology (QFort), Tainan, Taiwan.
[15]Physics Division, National Center for Theoretical Sciences, Taipei, Taiwan.
[16]Princeton Institute for Science and Technology of Materials, Princeton University, Princeton, New Jersey, USA.
[17]Lawrence Berkeley National Laboratory, Berkeley, California, USA.

These authors contributed equally: Xian P. Yang, Yueh-Ting Yao, Pengyu Zheng, Shuyue Guan.
Correspondence and requests for materials should be addressed to X.P.Y (email: xiany@princeton.edu) or to T.-R.C. (email: u32trc00@phys.nckn.edu.tw) or to M.Z.H. (email: mzhasan@princeton.edu)

**Abstract**
**The interplay of topology, magnetism, and correlations gives rise to intriguing phases of matter. In this study, through state-of-the-art angle-resolved photoemission spectroscopy, density functional theory and dynamical mean-field theory calculations, we visualize a fourfold degenerate Dirac nodal line at the boundary of the bulk Brillouin zone in the antiferromagnet $YMn_2Ge_2$. We further demonstrate that this gapless, antiferromagnetic Dirac nodal line is enforced by the combination of magnetism, space-time inversion symmetry and nonsymmorphic lattice symmetry. The corresponding drumhead surface states traverse the whole surface Brillouin zone. $YMn_2Ge_2$ thus serves as a platform to exhibit the interplay of multiple degenerate nodal physics and antiferromagnetism. Interestingly, the magnetic nodal line displays a *d*-orbital dependent renormalization along its trajectory in momentum space, thereby manifesting Hund's coupling. Our findings offer insights into the effect of electronic correlations on magnetic Dirac nodal lines, leading to an antiferromagnetic Hund nodal line.**

**Introduction**
Topological semimetals with protected band crossings have attracted intense interests recently [1-25]. Still, the combination of magnetism, spin-orbit coupling (SOC), and topology in real materials is quite uncommon. While there are fourfold degenerate Dirac nodal lines in some nonmagnetic materials [13-16], gapless fourfold AFM nodal lines have not been directly band-resolved experimentally. At the same time, correlated topological materials are beginning to emerge as the new frontier [26-32]. The interplay of non-trivial topology and electronic correlations can give rise to exotic phenomena such as superconductivity, fractionalization, and charge density waves [31, 32]. Despite the growing interest in this field, such quantum materials remain rare, particularly compared to their weakly correlated counterparts. Thus, there is a strong demand to uncover correlated topological magnets, where correlations and crystalline symmetry give rise to gapless topological states [27, 30].

In this work, we report an unusual gapless, fourfold degenerate AFM Dirac nodal line [21, 33] in $YMn_2Ge_2$ by using angle-resolved photoemission spectroscopy (ARPES) and first-principles calculations. We reveal that the Dirac nodal line in $YMn_2Ge_2$ is enforced to exhibit at the boundary lines of the Brillouin zone (BZ) by the combination of magnetism, space-time inversion symmetry and nonsymmorphic lattice symmetry. This specific symmetry protection allows $YMn_2Ge_2$ to present the largest magnetic Dirac nodal line in any real condensed matter system so far. We also highlight the observation of its bulk-boundary correspondence leading to remarkably long drumhead surface states traversing the length of the BZ. In addition, the AFM nodal line in $YMn_2Ge_2$ manifests a significant electronic correlation, as ARPES data are well described by theoretical calculations only after the inclusion of a non-negligible Hund's coupling [34, 35]. Although previous work [8, 30] discusses topological phases under correlations theoretically, experimental band-resolved demonstration is rare. In $YMn_2Ge_2$, the Dirac nodal line and the non-trivial drumhead state remain robust against strong electronic correlations. Our results demonstrate the effect of electronic correlations on non-trivial topology, giving rise to a Hund nodal line magnet [27]. Since the Dirac nodal line is guaranteed by symmetry, systems with similar magnetic or lattice symmetries can host similar correlated topological semimetallic phases.

**Results**
**Characterization of the fourfold degenerate AFM Dirac nodal line.** $YMn_2Ge_2$ crystallizes in the tetragonal structure with space group $I4/mmm$ (No. 139). As shown in Fig. 1**a**, $YMn_2Ge_2$ consists of Y-Ge-Mn-Ge-Y atomic layers stacked along the $c$ axis in a unit cell. Early magnetic measurements demonstrate an AFM-z ground state with a Néel temperature above the room temperature [36-39]. There are FM Mn atoms within each Mn layer and adjacent Mn layers are coupled antiferromagnetically (Fig. 1**a**). $YMn_2Ge_2$ hosts an AFM order, so it no longer exhibits ordinary time reversal symmetry $\mathcal{T}$. However, $YMn_2Ge_2$ maintains invariance under a combined symmetry $\tilde{\mathcal{T}} \equiv \{\mathcal{T} | \frac{1}{2}, \frac{1}{2}, \frac{1}{2}\}$. Our detailed symmetry analysis in the methods section shows how AFM fourfold degenerate nodal lines are protected by the combination of space-time inversion symmetry and nonsymmorphic lattice symmetry in $YMn_2Ge_2$ [21, 33]. Specifically, the joint $\mathcal{P}\tilde{\mathcal{T}}$ symmetry ensures globally doubly degenerate electronic bands in the whole BZ. When two such doubly degenerate bands cross on the boundary lines of the BZ with the additional protection of three glide mirror symmetries $\tilde{\mathcal{M}}_x \equiv \{\mathcal{M}_x | 0, \frac{1}{2}, \frac{1}{2}\}$, $\tilde{\mathcal{M}}_y \equiv \{\mathcal{M}_y | \frac{1}{2}, 0, \frac{1}{2}\}$, and $\tilde{\mathcal{M}}_z \equiv \{\mathcal{M}_z | \frac{1}{2}, \frac{1}{2}, 0\}$, fourfold degenerate AFM nodal line can be protected [33]. Breaking the nonsymmorphic mirror while maintaining the joint $\mathcal{P}\tilde{\mathcal{T}}$ symmetry in turn gaps out the nodal line (Supplementary Fig. 1). Interestingly, such nodal lines exist along the boundary of the 3D BZ, leading to a rare example of a maximized topological nodal line shown in Fig. 1**b**. As a result, the corresponding drumhead surface state could span across the length of the whole BZ.

To experimentally probe the band dispersion in $YMn_2Ge_2$, we first conduct a photon energy dependence scan along the $k_z$ direction and extract the photon energies corresponding to the Z-R-A high symmetry plane (Supplementary Fig. 2). The ARPES Fermi surface spectrum on the Z-R-A plane in Fig. 1**c** displays a clear fourfold rotational symmetry along the $k_z$ axis in $YMn_2Ge_2$, and our DFT calculations in Fig. 1**d** qualitatively match the experimental data. We then confirm the nodal line on the Z-R-A high symmetry plane using three cuts in Fig. 1**c**. The renormalized DFT calculation outlines the fourfold degenerate points at the A and R points (Figs. 1**f** and 1**h**). Moreover, Fig. 1**j** shows that the bands between the A and R points are also fourfold degenerate, which form a nodal line. ARPES data confirm this in cut 3 (Fig. 1**i**), and qualitatively agree with DFT calculations.

Having verified the nodal line, we now demonstrate its fourfold degeneracy. Due to joint $\mathcal{P}\tilde{\mathcal{T}}$ symmetry, every band should be doubly degenerate along the cut 1 direction. At the high symmetry A point, ARPES indicates that two doubly degenerate bands merge to form a fourfold degenerate point (Fig. 1**e**). Similarly, two doubly degenerate bands cross at a single point in cut 2, leading to the Dirac crossing at the R point (Fig. 1**g**). Additionally, bulk sensitive soft X-ray ARPES also reveal the nodal line along A-$\bar{\text{R}}$ direction and the Dirac cone at R point (Supplementary Fig. 3), confirming their bulk origin. It is intriguing to note that to better match ARPES data, all bulk electronic bands in DFT calculations are renormalized by a factor of 3, suggesting the presence of strong electronic correlations in YMn$_2$Ge$_2$. Despite the overall good agreement especially the energy position of the nodal point at the R point in Figs. 1**g-j** and the shape of the nodal line in Figs. 1**i-j**, the binding energy position of the Dirac crossing at the A point differs from renormalized DFT calculations, as shown by the red dashed line in Figs. 1**e-f**. Instead, a smaller renormalization factor (less than 3) should be applied at the A point so that the bands are less "compressed" below the Fermi level. As we will show later, this magnetic nodal line manifests a strong *d* orbital dependent renormalization due to Hund's coupling. Therefore, DFT results in Fig. 1 don't fully agree with ARPES data, indicating the necessity of DFT + DMFT calculations that will improve the agreement (Supplementary Fig. 4).

To obtain a comprehensive understanding of the fourfold AFM nodal line on the Z-R-A high symmetry plane, we utilized high-resolution photoemission measurements to trace the Dirac crossings along the A-R direction, as shown in Fig. 2. We have selected nine valence band cuts perpendicular to the nodal line (Fig. 2**b**) in Fig. 2**a**, which reveal the Dirac crossings displayed in Figs. 2**c-k**. To highlight the bulk nodal line dispersions clearly, we choose to show the second BZ in Fig. 2, since the surface states near the bulk bands are suppressed due to matrix element effect compared with the first BZ. By extracting the corresponding binding energy positions, as shown in Fig. 2**b**, the Dirac crossing moves to higher binding energies as the cuts move gradually away from the R point. The energy and momentum positions of these Dirac crossings are captured in Fig. 2**b**, which perfectly overlap with the nodal line, further confirming its fourfold degeneracy. Thus, the ARPES data offer a clear visualization of this AFM nodal line on the Z-R-A high symmetry plane. We also trace the nodal line along the A-M direction. As shown by Supplementary Figs. 5**a-f,** the fourfold crossing at $\bar{\text{M}}$ point remains gapless along the $k_z$ direction. Unlike the linear Dirac cone at the R point, the Dirac crossing at A point displays a parabolic dispersion (Supplementary Fig. 6). Moreover, while the nodal line along the A-R direction is gapless with SOC and magnetism included in Fig. 2, the bands are split along the M-X direction (Supplementary Figs. 5**g-i**). This verifies that only the boundary of the BZ holds fourfold degenerate nodes, eliminating the possibility of a gapless nodal surface on the whole A-R-X-M plane when SOC and magnetism are considered.

**Effect of the Hund's coupling on the AFM nodal line.** As we point out earlier, band dispersions at A point in DFT calculations deviate from ARPES data in Figs. 1**e-j**. This motivates us to check the Mn *d* orbital composition of YMn$_2$Ge$_2$, given that the electronic bands near the Fermi level are dominated by Mn *d* orbits. As shown in Fig. 1**j**, A and R points consist of mostly $d_{z^2}$ and $d_{xy}$ orbitals, respectively. Since distinct renormalization factors are needed for both points, an orbital dependent mass enhancement occurs for the nodal line in YMn$_2$Ge$_2$, which is an essential signature of a Hund's metal [40, 41]. In YMn$_2$Ge$_2$ under the AFM configuration, the more correlated $d_{xy}$ orbital is estimated to have a factor of 3 mass enhancement compared to DFT calculations, while the magnitude of m$^*$/m$_\text{DFT}$ for $d_{z^2}$ orbital is around 2. This strongly suggests that YMn$_2$Ge$_2$ is a Hund's metal.

We then compare our unrenormalized DFT (Fig. 3**a**) with DFT + DMFT calculations in Fig. 3**b**. In the resulting DFT+DMFT calculations, bands near the Fermi level are clearly renormalized compared to those obtained from DFT. Since the electronic bands away from the Fermi level are blurry, the Mn 3*d* electrons are expected to be incoherent due to electronic correlation effects. Therefore, electronic correlation is found to be a key factor that significantly impacts the electronic band structure, including the nodal line, in YMn$_2$Ge$_2$. Interestingly, the AFM nodal line and Dirac crossing persist in the DFT+DMFT results, as shown in Figs. 3**c** and 3**e**. More importantly, compared with those from DFT in Fig. 1**j** that already includes electron-electron correlations with a global 3 renormalization factor, the binding energy positions of the nodal points at R and A demonstrate a much better agreement with ARPES results when Hund's coupling is further considered (Figs. 3**d-e**). Specifically, the Dirac crossing at A residing at binding energy of 0.25 eV in Fig. 1**f** is pushed down by about 150 meV in Figs. 3**d-e**. This highlights the orbital dependent renormalization effect of the Hund's coupling on the AFM nodal line, in addition to the global renormalization already captured by DFT calculations. Furthermore, by adjusting the value of the Hund's coupling term (J) in DFT+DMFT, we confirm the presence of Hundness in YMn$_2$Ge$_2$ (Supplementary Fig. 7 and Supplementary Table 1). Our calculations reveal that the electronic band structure and the fluctuating moment in the paramagnetic state are highly sensitive to the magnitude of the Hund's coupling term (J). The spectral function undergoes a coherent-to-incoherent transition (Supplementary Fig. 7), with a large variation of *d*-orbital dependent quasi-particle mass enhancement and scattering rate (Supplementary Table 1). Moreover, the fluctuating moment in the paramagnetic state is calculated to increase from 2.62 $\mu_B$ for J=0.1 eV to 4.16 $\mu_B$ for J=0.8 eV, further indicating the presence of Hundness in YMn$_2$Ge$_2$.

**Drumhead surface states of the AFM nodal line.** After establishing the Hund-driven, orbital dependent renormalization of the bulk nodal line in YMn$_2$Ge$_2$, we investigate the associated drumhead surface states. They have been extensively studied in weakly interacting topological materials, but their counterparts in correlated semimetals are seldom explored [30]. Since the nodal line in YMn$_2$Ge$_2$ is along the boundary of the BZ on the A-R-Z plane, the drumhead surface state can span across the length of the whole surface BZ. Indeed, our calculations show the drumhead states connecting the Dirac crossing on the edge of the surface BZ (black arrows in Fig. 4**a**). We carefully trace their trajectory in the momentum space in Supplementary Fig. 8, which clearly displays two surface states emerging from the Dirac point. Specifically, they pass through the Fermi level and go below it again near the center of the BZ in Supplementary Fig. 8. Similarly, in our ARPES data (Fig. 4**b**), the surface states from the Dirac crossing at the boundary of the BZ are marked by the black dashed line. Moving towards the center of the BZ, they pass above the Fermi level and then move below the Fermi level, as shown in Figs. 4**b-c**. While there are three electron pockets near the center of the BZ in Fig. 4**b**, the middle one (highlighted by the red arrows) is the bulk state. At the same time, the outer two pockets (marked by the black arrows) are the drumhead surface states in Fig. 4**a**. Their photon energy dependence in Supplementary Fig. 9 reveals no dispersion along the $k_z$ direction, confirming their surface origin. Therefore, we demonstrate the exotically long drumhead surface state associated with the AFM nodal line based on ARPES data and calculations. We also extract and trace the energy dispersions of the bulk and surface states in Fig. 4**c**, which highlight the non-trivial bulk-boundary correspondence in YMn$_2$Ge$_2$. Remarkably, the surface states form an unusual dispersive three-dimensional network within the nodal line (Fig. 4**d**). A simplified way to understand them is to consider two twofold degenerate Weyl points that move towards each other until they form a fourfold degenerate Dirac crossing. Since a fermi arc is expected for each Weyl node, the two Fermi arc surface states from the original two Weyl cones could lead to the split drumhead surface states observed in ARPES, as further verified by the (001) surface constant energy contour calculations in Supplementary Fig. 10. Interestingly, these drumhead surface states show opposite spin polarization (Supplementary Fig. 11).

**Discussion**
Our findings demonstrate how Hund coupling can lead to an orbital dependent renormalization of the AFM nodal line quantum state in topological materials. Despite being topologically protected by the joint $\mathcal{P}\tilde{\mathcal{T}}$ symmetry plus nonsymmorphic lattice symmetry, a portion of the nodal line is shifted away from the Fermi level due to Hund's coupling, in addition to the global renormalization from electron-electron correlation. As a result, we observe the emergence of a renormalized AFM nodal line, manifesting the exotic interplay of nontrivial topology and strong electronic interactions [27]. In general, magnetic topological materials host tunable topological phases that can be controlled by various magnetic orders. YMn$_2$Ge$_2$ has the ThCr$_2$Si$_2$ type tetragonal structure, and the magnetic Mn sublattices display a vast variety of magnetic structures [42]. As shown in Supplementary Figs. 14 and 15, the AFM nodal line in YMn$_2$Ge$_2$ serves as a parent state leading to different magnetic topological phases including FM Weyl nodal surface (SmMn$_2$Ge$_2$ [43]) and AFM Dirac semimetal (TbMn$_2$Ge$_2$ [44]), which could reveal novel quantum responses such as third order current transport from the AFM nodal line [33] and AFM Dirac nodes induced anomalous skew-scattering nonlinear Hall effect [45]. As a result, the corresponding drumhead surface state can be tuned with various magnetic phases. More importantly, since Mn *d* orbitals manifest significant Hund's coupling, how Hund's coupling interacts with various magnetic topological phases in RMn$_2$Ge$_2$ compounds needs more studies in the future.

**Methods**

**Single crystal synthesis.** Single crystals of $YMn_2Ge_2$ were synthesized using the flux method. The initial atomic ratio of Y : Mn : Ge was 1 : 10 : 10, with additional Mn and Ge used as fluxes. The mixture was placed in an alumina crucible and sealed in evacuated quartz ampoules under vacuum. The ampoule was heated to 1050 °C over 5 hours and kept for 24 hours. It was then cooled to 1000 °C at the rate of 5 °C/h and heated again to 1040 °C to decrease the number of seed crystals. Finally, it was slowly cooled to 900 °C at the rate of 3 °C/h. After centrifugation, several flat millimeter-sized crystals were obtained.

**Angle-resolved photoemission spectroscopy.** Vacuum ultraviolet ARPES measurements (VUV-ARPES) were conducted out at Beamline 5-2 of the Stanford Synchrotron Radiation Lightsource (SSRL) at SLAC in Menlo Park, CA, USA and Bloch beamline in the MAX IV Laboratory in Lund, Sweden. The energy resolution was better than 20 meV and temperature was less than 20 K. Samples were cleaved in situ and measured under vacuum better than $5 \times 10^{-11}$ Torr. The Soft X-ray ARPES data were obtained at BL25SU of Spring-8 in Japan, with an energy resolution of around 100 meV. Preliminary ARPES data were also collected at beamline 4.0.3 at the Advanced Light Source (ALS) in Berkeley, CA, USA and Beamline 21-ID-1 (ESM-ARPES) of the National Synchrotron Light Source II, BNL in Upton, NY, USA.

ARPES dispersion map in Fig. 4**b** was probed with 148 eV photon energy. The high-resolution dispersion map embedded at the center was collected with 97 eV incident light. Figs. 1**e** and 1**g** were taken with 97 eV photon energy. Supplementary Fig. 3 and Supplementary Fig. 5**g** were taken with Soft X-ray light of 552eV and 500 eV, respectively. The rest of ARPES data in the main text were probed with 148 eV light. Photon energy dependence in Supplementary Fig. 2 was taken with energies from 60 to 200 eV. All ARPES data were collected with LH polarization. The photon energy scan in Supplementary Fig. 2 demonstrates a clear periodic pattern along the $k_z$ direction. An inner potential of 11 eV was accordingly extracted. Therefore, photon energies of 148 and 97 eV probe the Z-A-R plane. Thus, the ARPES data presented in the main text correspond to the Z-A-R high symmetry plane. The experimentally observed termination is the Ge termination after a careful comparison between ARPES spectrum and theoretical calculations (Figs. 1 and 4).

**First-principles calculations.** The bulk band structures of $YMn_2Ge_2$ were computed using the projector augmented wave method as implemented in the VASP package [46-48] within the GGA [49] scheme. The experimental lattice parameters and the magnetic configuration were used [50]. The spin-orbit coupling (SOC) was included self consistently in the calculations of electronic structures with a Monkhorst-Pack k-point mesh $15 \times 15 \times 5$. We constructed a tight-binding Hamiltonian for $YMn_2Ge_2$, where the tight-binding model matrix elements were calculated by projecting onto the Wannier orbitals, which used the VASP2WANNIER90 interface [51-53]. The Y *d* orbitals, Mn *s*, and *d* orbitals, and Ge *p* orbitals were used to construct the Wannier functions without performing the maximizing localization. The surface state electronic structure was calculated by the surface Green's function technique, which computes the spectral weight near the surface of a semi-infinite system.

**DFT+DMFT calculations.** To take into account of the electronic correlation effect, we performed density functional theory plus dynamical mean-field theory (DFT+DMFT) calculations [54,55] to compute the electronic structure of $YMn_2Ge_2$ in the antiferromagnetic state with SOC included. The DFT part was calculated by the linearized augmented plane-wave method as implemented in WIEN2K [56]. The PBE exchange-correlation functional was used. We treated Mn 3*d* orbitals correlated and Hubbard U=5 eV, Hund's coupling J=0.8 eV were used, in consistent with the previous calculations [57-60]. We chose the "exact" double-counting scheme to solve the double counting problem [61]. The impurity problem in DFT+DMFT was solved by using the continuous time quantum Monte Carlo method [62,63] at T=100K. For Figs. 3**c** and **e**, we shifted the Fermi level to match ARPES results and reduced the imaginary part of the self-energy when we plotted by a factor of 0.5 to make them clearer. We used the experimental lattice parameters in all our calculations [50].

To firmly confirm that $YMn_2Ge_2$ is a Hund's metal, we also took DFT+DMFT calculations in the paramagnetic states without considering SOC for different Hund's coupling settings (J=0.8 eV and J=0.1 eV), and other settings were consistent with the above calculations. We present our results in Supplementary Fig. 7 and Supplementary

Table 1. We obtained the quasi-particle mass enhancement $\frac{m^*}{m_{band}} = 1 - \frac{\partial \text{Im}\Sigma(i\omega_n)}{\partial \omega_n}|_{\omega_n \to 0^+}$ and scattering rate $\Gamma = -\text{Im}\Sigma(i0^+)$ by the three-order polynominal fit of the self energy at the first five Matsubara frequencies.

**Symmetry analysis of YMn$_2$Ge$_2$.** We will show below that YMn$_2$Ge$_2$ with paramagnetic phase possesses the Dirac nodal surface band dispersion. The fourfold degenerate band dispersion occurs at $k_{i=x,y,z} = \pi$ (R-Z-A) plane (Supplementary Fig. 14**c**), which is protected by space-time inversion symmetry, mirror symmetry and glide mirror symmetry. After considering AFM spin configuration, mirror symmetry is broken. Consequently, the band splits from fourfold degeneracy to twofold degeneracy, except at the boundary lines of the BZ (Supplementary Fig. 14**a**). As a result, the inclusion of AFM leads to the Dirac nodal lines at the boundary of the BZ observed in experiment. Additionally, when rotating the spin configuration from AFM to FM (we assume the direction of spin is along z-axis), the space-time inversion symmetry and the mirror symmetry $\mathcal{M}_x$ and $\mathcal{M}_y$ are further broken. Therefore, the band structure of YMn$_2$Ge$_2$ with FM-z shows singlet degeneracy at generic $k$ points in the BZ, while preserves doubly degenerate band dispersion on the mirror plane $\mathcal{M}_z$ (Supplementary Fig. 14**b**), leading to a FM Weyl nodal surface on the R-Z-A plane. The detail is below.

*Paramagnetic phase*: The space group of YMn$_2$Ge$_2$ with paramagnetic phase is I4/mmm (#139). This space group has spatial-inversion symmetry $\mathcal{P}$. Since time-reversal symmetry $\mathcal{T}$ is preserved, the anti-unitary symmetry $\mathcal{PT}$ is preserved at arbitrary $k$ point and satisfies $(\mathcal{PT})^2 = -1$, which leads to the double-degeneracy of the bands at any generic momentum $k$. The space group I4/mmm contains two alternative origins, (0, 0, 0) and (1/2, 1/2, 1/2), resulting in a half-translation symmetry $t \equiv \{1|\frac{1}{2},\frac{1}{2},\frac{1}{2}\}$. Therefore, both the mirror symmetry $\mathcal{M}_z$ and the glide mirror symmetry $\widetilde{\mathcal{M}}_z \equiv \{\mathcal{M}_z|\frac{1}{2},\frac{1}{2},\frac{1}{2}\}$ are preserved on the $k_z = 0$ and $\pi$. Let $\widetilde{\mathcal{M}}_z$ and $\mathcal{PT}$ act on $(x, y, z)$ in different orders,

$$(x, y, z) \xrightarrow{\mathcal{PT}} (-x, -y, -z) \xrightarrow{\widetilde{\mathcal{M}}_z} \left(-x + \frac{1}{2}, -y + \frac{1}{2}, z + \frac{1}{2}\right) \quad (1)$$

$$(x, y, z) \xrightarrow{\widetilde{\mathcal{M}}_z} \left(x + \frac{1}{2}, y + \frac{1}{2}, -z + \frac{1}{2}\right) \xrightarrow{\mathcal{PT}} \left(-x - \frac{1}{2}, -y - \frac{1}{2}, z - \frac{1}{2}\right) \quad (2)$$

We can derive the following relation on the $k_z = \pi$, $\widetilde{\mathcal{M}}_z(\mathcal{PT}) = -e^{-i(k_x+k_y)}(\mathcal{PT})\widetilde{\mathcal{M}}_z$. In addition, $\widetilde{\mathcal{M}}_z$ is anticommute with $\mathcal{M}_z$ on the $k_z = \pi$, which can be obtained by applying $\widetilde{\mathcal{M}}_z$ and $\mathcal{M}_z$ on $(x, y, z)$ in the reverse order,

$$(x, y, z) \xrightarrow{\mathcal{M}_z} (x, y, -z) \xrightarrow{\widetilde{\mathcal{M}}_z} \left(x + \frac{1}{2}, y + \frac{1}{2}, z - \frac{1}{2}\right) \quad (3)$$

$$(x, y, z) \xrightarrow{\widetilde{\mathcal{M}}_z} \left(x + \frac{1}{2}, y + \frac{1}{2}, -z + \frac{1}{2}\right) \xrightarrow{\mathcal{M}_z} \left(x + \frac{1}{2}, y + \frac{1}{2}, z - \frac{1}{2}\right) \quad (4)$$

Then we obtain $\widetilde{\mathcal{M}}_z \mathcal{M}_z = -\mathcal{M}_z \widetilde{\mathcal{M}}_z$ on the $k_z = \pi$. Since $(\widetilde{\mathcal{M}}_z)^2 = -e^{-i(k_x+k_y)}$ on $k_z = \pi$, we have the $\widetilde{\mathcal{M}}_z$ eigenvalue $g_z = \pm i e^{-i(k_x+k_y)/2}$. We consider an eigenstate $|\psi(\mathbf{k})\rangle$ of $\widetilde{\mathcal{M}}_z$ with the eigenvalue $g_z$ on the $k_z = \pi$, $\widetilde{\mathcal{M}}_z |\psi(\mathbf{k})\rangle = g_z |\psi(\mathbf{k})\rangle$. Because of $\{\widetilde{\mathcal{M}}_z, \mathcal{M}_z\} = 0$, we have $\widetilde{\mathcal{M}}_z(\mathcal{M}_z|\psi(\mathbf{k})\rangle) = -g_z(\mathcal{M}_z|\psi(\mathbf{k})\rangle)$. This result indicates band degeneracy throughout the whole surface of $k_z = \pi$, with the degenerate eigenstates carrying opposite eigenvalues for $\widetilde{\mathcal{M}}_z$ and $\mathcal{M}_z$. In the next, we examine the $\widetilde{\mathcal{M}}_z$ eigenvalues of the two Kramer's degenerate states that are related by $\mathcal{PT}$. We select two degenerate eigenstates on the $k_z = \pi$, $|\psi(\mathbf{k})\rangle$ and $\mathcal{PT}|\psi(\mathbf{k})\rangle$, in which $\widetilde{\mathcal{M}}_z|\psi(\mathbf{k})\rangle = g_z|\psi(\mathbf{k})\rangle$, where $g_z = ie^{-i(k_x+k_y)/2}$. The relation between the $\widetilde{\mathcal{M}}_z$ eigenvalues of these states is below,

$$\widetilde{\mathcal{M}}_z(\mathcal{PT}|\psi(\mathbf{k})\rangle) = -e^{-i(k_x+k_y)}(\mathcal{PT})\widetilde{\mathcal{M}}_z|\psi(\mathbf{k})\rangle \quad (5)$$
$$= -e^{-i(k_x+k_y)}(\mathcal{PT})g_z|\psi(\mathbf{k})\rangle$$

$$= -e^{-i(k_x+k_y)}g_z^*(\mathcal{PT})|\psi(\mathbf{k})\rangle$$
$$= g_z(\mathcal{PT}|\psi(\mathbf{k})\rangle)$$

This indicates that the degenerate eigenstates of Kramer's states, linked through $\mathcal{PT}$, exhibit an identical $\widetilde{\mathcal{M}}_z$ eigenvalue $g_z$. Consequently, an additional pair of Kramer's states becomes necessary, characterized by the $\widetilde{\mathcal{M}}_z$ eigenvalue of $-g_z$. Therefore, in total the four states ($|\psi(\mathbf{k})\rangle$, $\mathcal{PT}|\psi(\mathbf{k})\rangle$, $\mathcal{M}_z|\psi(\mathbf{k})\rangle$, $\mathcal{PTM}_z|\psi(\mathbf{k})\rangle$) form a degenerate quartet on the $k_z = \pi$. This conclusion remains valid for $k_x = \pi$ and $k_y = \pi$.

Thus, YMn$_2$Ge$_2$ with paramagnetic phase hosts the Dirac nodal surface at $k_{i=x,y,z} = \pi$ plane, which is protected by space-time inversion symmetry, mirror symmetry and glide mirror symmetry.

*Antiferromagnetic phase*: According to the DFT calculations (Fig. 1 and Supplementary Fig. 1), the fourfold degenerate Dirac nodal line exists at the boundary line $(k_x, k_y, k_z) = (\pi, k, \pi)$, $(k, \pi, \pi)$, and $(\pi, \pi, k)$ of the Brillouin zone (BZ). We now show that these band degeneracies are protected by the combination of the space-time inversion symmetry and nonsymmorphic lattice symmetry. YMn$_2$Ge$_2$ hosts antiferromagnetic (AFM) order, thus the system no longer exhibits ordinary time reversal symmetry $\mathcal{T}$. However, it maintains invariance under a combined symmetry $\tilde{\mathcal{T}} \equiv \left\{\mathcal{T}\middle|\frac{1}{2},\frac{1}{2},\frac{1}{2}\right\}$. Taking the spin space into account, $\tilde{\mathcal{T}}$ can be represented as $\tilde{\mathcal{T}} = e^{-i(k_x+k_y+k_z)/2}i\sigma_y\mathcal{K}$, where $\mathcal{K}$ stands for complex conjugation and $\sigma_{i=x,y,z}$ denote the Pauli matrices for spin. The magnetic space group of YMn$_2$Ge$_2$ (#126) preserves spatial inversion symmetry $\mathcal{P}$. Therefore, the anti-unitary symmetry $\mathcal{P}\tilde{\mathcal{T}}$ is preserved at arbitrary $k$ point and satisfies $(\mathcal{P}\tilde{\mathcal{T}})^2 = -1$. Consequently, all bands are doubly degenerate at every generic $k$ in the whole BZ. Next, we consider $\mathcal{P}\tilde{\mathcal{T}}$ symmetry and three glide mirror symmetries $\widetilde{\mathcal{M}}_x \equiv \left\{\mathcal{M}_x\middle|0,\frac{1}{2},\frac{1}{2}\right\}$, $\widetilde{\mathcal{M}}_y \equiv \left\{\mathcal{M}_y\middle|\frac{1}{2},0,\frac{1}{2}\right\}$, and $\widetilde{\mathcal{M}}_z \equiv \left\{\mathcal{M}_z\middle|\frac{1}{2},\frac{1}{2},0\right\}$ preserved respectively on the boundary lines of the BZ to demonstrate the symmetry enforced fourfold degeneracy of nodal lines. We first concentrate on the line $k = (\pi, k, \pi)$. Let $\widetilde{\mathcal{M}}_z$ and $\mathcal{P}\tilde{\mathcal{T}}$ act on $(x, y, z)$ in the different orders,

$$(x,y,z) \xrightarrow{\mathcal{P}\tilde{\mathcal{T}}} \left(-x-\tfrac{1}{2},-y-\tfrac{1}{2},-z-\tfrac{1}{2}\right) \xrightarrow{\widetilde{\mathcal{M}}_z} \left(-x,-y,z+\tfrac{1}{2}\right) \tag{6}$$

$$(x,y,z) \xrightarrow{\widetilde{\mathcal{M}}_z} \left(x+\tfrac{1}{2},y+\tfrac{1}{2},-z\right) \xrightarrow{\mathcal{P}\tilde{\mathcal{T}}} \left(-x-1,-y-1,z-\tfrac{1}{2}\right) \tag{7}$$

These two combinations are connected by the momentum-dependent phase $\widetilde{\mathcal{M}}_z(\mathcal{P}\tilde{\mathcal{T}}) = e^{-ik_y}(\mathcal{P}\tilde{\mathcal{T}})\widetilde{\mathcal{M}}_z$. Since $(\widetilde{\mathcal{M}}_z)^2 = e^{-ik_y}$ at $(\pi, k, \pi)$, we have the $\widetilde{\mathcal{M}}_z$ eigenvalue $g_z = \pm e^{-ik_y/2}$. We consider an eigenstate $|\psi(\mathbf{k})\rangle$ of $\widetilde{\mathcal{M}}_z$ with the eigenvalue $g_z$ at $k = (\pi, k, \pi)$, $\widetilde{\mathcal{M}}_z|\psi(\mathbf{k})\rangle = g_z|\psi(\mathbf{k})\rangle$. Then we show that its Kramer partner $\mathcal{P}\tilde{\mathcal{T}}|\psi(\mathbf{k})\rangle$ is the eigenstate of $\widetilde{\mathcal{M}}_z$, which share the same eigenvalue $g_z$,

$$\widetilde{\mathcal{M}}_z(\mathcal{P}\tilde{\mathcal{T}}|\psi(\mathbf{k})\rangle) = e^{-ik_y}(\mathcal{P}\tilde{\mathcal{T}})\widetilde{\mathcal{M}}_z|\psi(\mathbf{k})\rangle \tag{8}$$
$$= e^{-ik_y}(\mathcal{P}\tilde{\mathcal{T}})g_z|\psi(\mathbf{k})\rangle$$
$$= e^{-ik_y}g_z^*(\mathcal{P}\tilde{\mathcal{T}})|\psi(\mathbf{k})\rangle$$
$$= g_z(\mathcal{P}\tilde{\mathcal{T}}|\psi(\mathbf{k})\rangle)$$

However, $\widetilde{\mathcal{M}}_z$ is anticommute with $\widetilde{\mathcal{M}}_x$ along the line $(\pi, k, \pi)$, which can be obtained by applying $\widetilde{\mathcal{M}}_z$ and $\widetilde{\mathcal{M}}_x$ on $(x, y, z)$ in the reverse orders,

$$(x,y,z) \xrightarrow{\widetilde{\mathcal{M}}_x} \left(-x, y+\tfrac{1}{2}, z+\tfrac{1}{2}\right) \xrightarrow{\widetilde{\mathcal{M}}_z} \left(-x+\tfrac{1}{2}, y+1, -z-\tfrac{1}{2}\right) \tag{9}$$

$$(x,y,z) \xrightarrow{\widetilde{\mathcal{M}}_z} \left(x+\tfrac{1}{2}, y+\tfrac{1}{2}, -z\right) \xrightarrow{\widetilde{\mathcal{M}}_x} \left(-x-\tfrac{1}{2}, y+1, -z+\tfrac{1}{2}\right) \tag{10}$$

Thus, the result above shows that $\widetilde{\mathcal{M}}_z\widetilde{\mathcal{M}}_x = -\widetilde{\mathcal{M}}_x\widetilde{\mathcal{M}}_z$ along the line, where the minus sign originates from anticommutation relation of Pauli matrices for spin space. Because $\{\widetilde{\mathcal{M}}_z, \widetilde{\mathcal{M}}_x\} = 0$, we have $\widetilde{\mathcal{M}}_z(\widetilde{\mathcal{M}}_x|\psi(\mathbf{k})\rangle) = -g_z(\widetilde{\mathcal{M}}_x|\psi(\mathbf{k})\rangle)$. This indicates that the band degeneracy is required, and the involved bands should carry opposite $\widetilde{\mathcal{M}}_z$ eigenvalue $-g_z$. The fact that the eigenstates $|\psi(\mathbf{k})\rangle$ and $\mathcal{P}\widetilde{\mathcal{T}}|\psi(\mathbf{k})\rangle$ share the same eigenvalue $g_z$, indicates another pair of Kramer's states that carry $-g_z$ is required. Therefore, $\widetilde{\mathcal{M}}_x|\psi(\mathbf{k})\rangle$ as well as $\mathcal{P}\widetilde{\mathcal{T}}\widetilde{\mathcal{M}}_x|\psi(\mathbf{k})\rangle$ must be another degenerate partner of $|\psi(\mathbf{k})\rangle$ with the opposite eigenvalue $-g_z$. Thus, in total the four states ($|\psi(\mathbf{k})\rangle$, $\mathcal{P}\widetilde{\mathcal{T}}|\psi(\mathbf{k})\rangle$, $\widetilde{\mathcal{M}}_x|\psi(\mathbf{k})\rangle$, $\mathcal{P}\widetilde{\mathcal{T}}\widetilde{\mathcal{M}}_x|\psi(\mathbf{k})\rangle$) guarantee the formation of a degenerate quartet at the boundary line $(\pi, k, \pi)$. This conclusion holds true for other boundaries $(k, \pi, \pi)$ and $(\pi, \pi, k)$.

Therefore, the fourfold degenerate Dirac nodal line (instead of nodal surface) at the boundary line of BZ in AFM YMn$_2$Ge$_2$ is symmetry enforced by the combination of the space-time inversion symmetry and nonsymmorphic lattice symmetry. The corresponding surface states should also be topologically protected. We further add their photon energy dependence as Supplementary Fig. 9 to confirm the surface origin of the drumhead states.

**Quadratic band dispersions near the A point.** In addition to DFT calculations in Supplementary Fig. 6, the band dispersion from A to Z can be demonstrated theoretically. Since the bands along the boundaries of the BZ such as A-R and A-M are fourfold degenerate, we need to construct a 4-band model, constrained by the symmetries preserved. At A point, the symmetries can be generated by $\widetilde{\mathcal{T}}$, $\mathcal{P}$, $\widetilde{\mathcal{M}}_y$, and $C_{4z}$. Moreover, symmetries at A point satisfy the relationship $\{\mathcal{P}\widetilde{\mathcal{T}}, \widetilde{\mathcal{M}}_i\} = 0$, $\{\widetilde{\mathcal{M}}_i, \widetilde{\mathcal{M}}_j\} = -2\delta_{ij}$, $\{\mathcal{P}, C_{4z}\} = 0$, $(\mathcal{P}\widetilde{\mathcal{T}})^2 = -1$, and $\widetilde{\mathcal{T}}^2 = 1$. Below we prove $\{\mathcal{P}, C_{4z}\} = 0$. For under the operations of $\mathcal{P}$ and $C_{4z}$, we have

$$(x, y, z) \xrightarrow{C_{4z}} \left(-y + \frac{1}{2}, x, z\right) \xrightarrow{\mathcal{P}} \left(y - \frac{1}{2}, -x, -z\right) \tag{11}$$

$$(x, y, z) \xrightarrow{\mathcal{P}} (-x, -y, -z) \xrightarrow{C_{4z}} \left(y + \frac{1}{2}, -x, -z\right) \tag{12}$$

This indicates $C_{4z}\mathcal{P} = e^{-ik_y}\mathcal{P}C_{4z}$. As a result, $\mathcal{P}$ and $C_{4z}$ satisfy anticommutation relation $\{\mathcal{P}, C_{4z}\} = 0$ at A point.

According to the four anticommutation relation at A point, we select the constraints of the symmetries at A point as $\mathcal{P}\widetilde{\mathcal{T}} = \tau_1\sigma_2\mathcal{K}$, $\widetilde{\mathcal{T}} = i\tau_2\sigma_2\mathcal{K}$, $\mathcal{P} = \tau_3$, $\widetilde{\mathcal{M}}_y = i\tau_3\sigma_2$, and $C_{4z} = \tau_1 \otimes e^{-i\frac{\pi}{4}\sigma_3}$. Consequently, the $\mathbf{k}\cdot\mathbf{p}$ low-energy effective model is constructed with three $\Gamma$-matrices, $\tau_3\sigma_{1,2,3}$, which can be expressed as

$$H_A(\overline{\mathbf{q}}) = \alpha(q_x^2 + q_y^2)I_4 + \beta q_z^2 I_4 + \gamma q_x q_y \tau_3\sigma_3 + \lambda(q_z q_x \tau_3\sigma_2 - q_z q_y \tau_3\sigma_1) \tag{13}$$

Based on $H_A$, we find that the bands split into two doubly degenerate bands with parabolic dispersion when away from A point ($q_i = 0$).

**Zak phase of the drumhead surface states.** We provide a theoretical point of view to illustrate why drumhead surface states are expected in YMn$_2$Ge$_2$. The drumhead surface state in a nodal line semimetal is generally protected by a quantized $\pi$ Zak phase, which represents the Berry phase of a straight line normal to the surface and crossing the bulk BZ [64]. Supplementary Fig. 16 shows the Zak phase $\mathcal{Z}(k_x, k_y)$ along the $k_z$ axis in the bulk BZ, perpendicular to the (001) plane, by using Wilson loop [64]. Our calculations demonstrate that the Zak phase at point **a** is quantized to $-\pi$, denoted as $\mathcal{Z}(\mathbf{a}) = -\pi$ (Supplementary Fig. 16**a**). The Zak phase remains constant until the line intersects the nodal line. Due to symmetries, the Dirac nodal line in YMn$_2$Ge$_2$ is constrained to lie on the BZ boundary (blue lines), resulting in a quantized Zak phase $\mathcal{Z}(k_x, k_y) = -\pi$ at any point on the $k_x - k_y$ plane, as shown in Supplementary Figs. 16**b-d**. When the line crosses the Dirac nodal line at the BZ boundary, the Zak phase changes by $2\pi$, corresponding to the Berry phase of the Dirac nodal line. Consequently, the Zak phase $\mathcal{Z}(k_x, k_y) = -\pi$ for any gapped state leading to the formation of drumhead surface states on the (001) surface. Therefore, the appearance of drumhead states in YMn$_2$Ge$_2$ is protected by the bulk nodal line structure, and thus should be expected.

**Data Availability**. All relevant data are available from the corresponding authors upon request.

**Acknowledgments**
The authors thank Junyi Zhang, Qi Zhang and Jonathan Denlinger for the fruitful discussions. Advanced ARPES and theoretical work at Princeton University were supported by the US DOE under the Basic Energy Sciences program (grant number DOE/BES DE-FG-02-05ER46200; M.Z.H.), NQI at ORNL_Quantum Science Center supported by US DOE and the Gordon and Betty Moore Foundation (GBMF4547 and GBMF9461; M.Z.H.). Use of the Stanford Synchrotron Radiation Lightsource, SLAC National Accelerator Laboratory, is supported by the U.S. Department of Energy, Office of Science, Office of Basic Energy Sciences under Contract No. DE-AC02-76SF00515. This research used resources of the Advanced Light Source, which is a DOE Office of Science User Facility under contract no. DE-AC02-05CH11231. We acknowledge MAX IV Laboratory for time on Beamline Bloch under Proposal 20210268. Research conducted at MAX IV, a Swedish national user facility, is supported by the Swedish Research council under contract 2018-07152, the Swedish Governmental Agency for Innovation Systems under contract 2018-04969, and Formas under contract 2019-02496. Synchrotron radiation experiments were performed at the BL25SU of SPring-8 with the approval of the Japan Synchrotron Radiation Research Institute (JASRI) (Proposal No. 2023A1611). This research used the ESM beamline of the National Synchrotron Light Source II, a U.S. Department of Energy (DOE) Office of Science User Facility operated for the DOE Office of Science by Brookhaven National Laboratory under Contract No. DE-SC0012704. The authors want to thank D. Lu, M. Hashimoto at Beamline 5-2 of the Stanford Synchrotron Radiation Lightsource (SSRL) at the SLAC National Accelerator Laboratory, C. Polley, J. Adell and B. Thiagarajan at Beamline Bloch of the Max IV, and K. Yamagami at BL25SU of Spring-8 for support. The authors also thank J. Denlinger at Beamline 4.0.3 (MERLIN) of the ALS, E. Vescovo and T. Yilmaz at Beamline 21-ID-1 (ESM-ARPES) of the National Synchrotron Light Source II for support in getting the preliminary ARPES data. The work in Peking University was financially supported by the National Natural Science Foundation of China (Grants No. 12225401, No. 12141002, and No. 12274154), the National Key Research and Development Program of China (Grant No. 2021YFA1401902), the CAS Interdisciplinary Innovation Team, the Strategic Priority Research Program of Chinese Academy of Sciences (Grant No. XDB28000000). T.-R.C. was supported by the 2030 Cross-Generation Young Scholars Program from the National Science and Technology Council (NSTC) in Taiwan (Program No. MOST111-2628-M-006-003-MY3), National Cheng Kung University (NCKU), Taiwan, and National Center for Theoretical Sciences, Taiwan. T.-R.C. thanks to National Center for High-performance Computing (NCHC) of National Applied Research Laboratories (NARLabs) in Taiwan for providing computational and storage resources. This research was supported, in part, by Higher Education Sprout Project, Ministry of Education to the Headquarters of University Advancement at NCKU. Zhiping Yin and Pengyu Zheng were supported by Innovation Program for Quantum Science and Technology (2021ZD0302800), the Fundamental Research Funds for the Central Universities (Grant No. 2243300003), and the National Natural Science Foundation of China (Grant No. 12074041). The DFT + DMFT calculations were carried out with high-performance computing cluster of Beijing Normal University in Zhuhai. The experimental work at Huazhong University of Science and Technology was supported by the National Natural Science Foundation of China (Grant No. 11874158), and Guangdong Basic and Applied Basic Research Foundation (Grant No.2022B1515120020). The theory work at Huazhong University of Science and Technology was supported by the National Key Research and Development Program of China (2018YFA0307000), and the National Natural Science Foundation of China (12274154,11874022) T.A.C. acknowledges the support of the National Science Foundation Graduate Research Fellowship Program (DGE-1656466). The work at Northeastern University is supported by the Air Force Office of Scientific Research under award number FA9550-20-1-0322. I.B. acknowledges the generous support of the Special Postdoctoral Researchers Program, RIKEN during the late stages of this work. M.Z.H. acknowledges support from Lawrence Berkeley National Laboratory and the Miller Institute of Basic Research in Science at the University of California, Berkeley in the form of a Visiting Miller Professorship. M.Z.H. also acknowledges support from the U.S. Department of Energy, Office of Science, National Quantum Information Science Research Centers (ORNL), Quantum Science Center and Princeton University.

**Author Contributions.**
X.P.Y, T.-R.C. and M.Z.H conceived the project. ARPES experiments were performed out by X.P.Y. with close assistance from T.A.C., Z.-J.C. and I.B. in consultation with M.Z.H. Crystal growth was carried out by S.G., H.Z., Z.L. and T.S. in consultation with S.J. and Z.T. First-principal calculations were carried out by Y.-T.Y., P.Z. and C.L. with assistance from S.C. and G.X. in consultation with Z.Y. and T.-R.C. X.Z. performed symmetry analysis of the nodal line in consultation with A.B. M.S.H, Y.-X. J. and M.L. contributed to the calibration of the measurement. X.P.Y., Y.-T.Y. and P.Z. performed the analysis, interpretation, and figure development in consultation with J.-X.Y., Z.Y., T.-R.C. and M.Z.H. X.P.Y., Y.-T.Y., T.-R.C. and M.Z.H. wrote the manuscript with input from all other authors. T.-R.C. and M.Z.H. supervised the project.

**Competing Interests**.
The authors declare no competing interests.

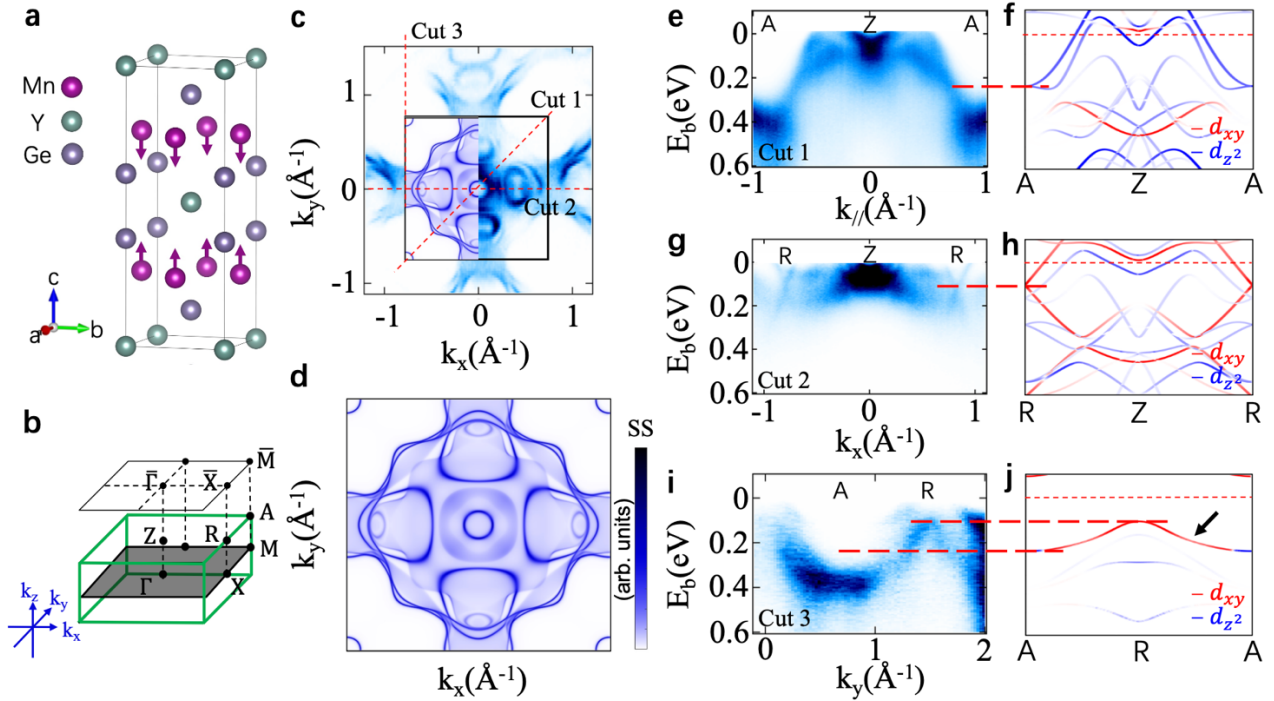

**Fig. 1. Fermi surface and AFM Dirac line in YMn$_2$Ge$_2$. a,** The crystal structure of YMn$_2$Ge$_2$. The purple arrows indicate that the two adjacent ferromagnetic Mn layers have opposite spin directions. **b,** Bulk and surface Brillouin zones (BZs) of YMn$_2$Ge$_2$. High symmetry points are marked. The magnetic nodal line around the boundary of the BZ is highlighted by the green line. **c,** ARPES Fermi surface spectrum on the (001) surface. The black square represents the surface BZ. Red dashed lines indicate ARPES dispersion cuts 1-3 in **e-j**. **d,** Calculated Fermi surface map corresponding to the black box in **c** and integrated over all the $k_z$ values. The same plot is also embedded in **c**. **e-f,** ARPES dispersion map (**e**) and the corresponding bulk band structure calculation (**f**) along cut 1 in **c**. SS stands for surface state. High symmetry points are marked. Mn $d_{z^2}$ (blue) and $d_{xy}$ (red) orbitals are projected on the bulk bands. The Fermi level in **f** is adjusted according to the experimental data and DFT is renormalized by 3. Two doubly degenerate bands cross at the A point to form a fourfold degeneracy. Thick red dashed line shows that the binding energy of the Dirac crossing at the A point in the renormalized DFT differs from experimental data. **g-h,** ARPES dispersion map (**g**) and the corresponding bulk band structure calculation (**h**) along cut 2 in **c**. A fourfold Dirac crossing can be seen at the R point. The thick red dashed line shows the consistency between the renormalized DFT and experimental data especially at the R point. **i-j,** ARPES dispersion map (**i**) and the corresponding bulk band structure calculation (**j**) along cut 3 in **c**. Thick red dashed lines suggest that although the renormalized DFT correctly describes the band dispersion near the R point, it doesn't agree well with ARPES data at the A point. Black arrow in **j** indicates the nodal line, and A and R points consist of mostly $d_{z^2}$ and $d_{xy}$ orbitals, respectively.

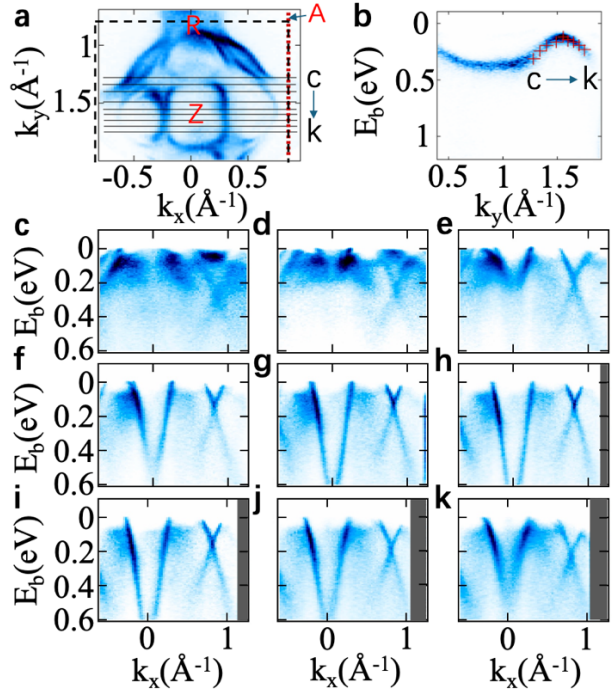

**Fig. 2. Detailed characterization of the AFM nodal line. a**, ARPES Fermi surface spectrum on the (001) surface in the second Brillouin zone. The black dotted box represents the second Brillouin zone and corresponding high symmetry points are marked. Red dashed line on the boundary of the Brillouin zone indicates the dispersion map in **b**. Nine black lines represent the dispersion maps from **c-k**. **b,** ARPES dispersion map along the A-R-A high symmetry direction. Red cross signs are the extracted energy and momentum positions of the Dirac crossings in **c-k**. **c-k,** Corresponding ARPES dispersion maps demonstrating the fourfold degenerate crossings along the AFM nodal line. The grey-shaded areas in **h-k** mask the distorted spectra at large angles of the ARPES analyzer.

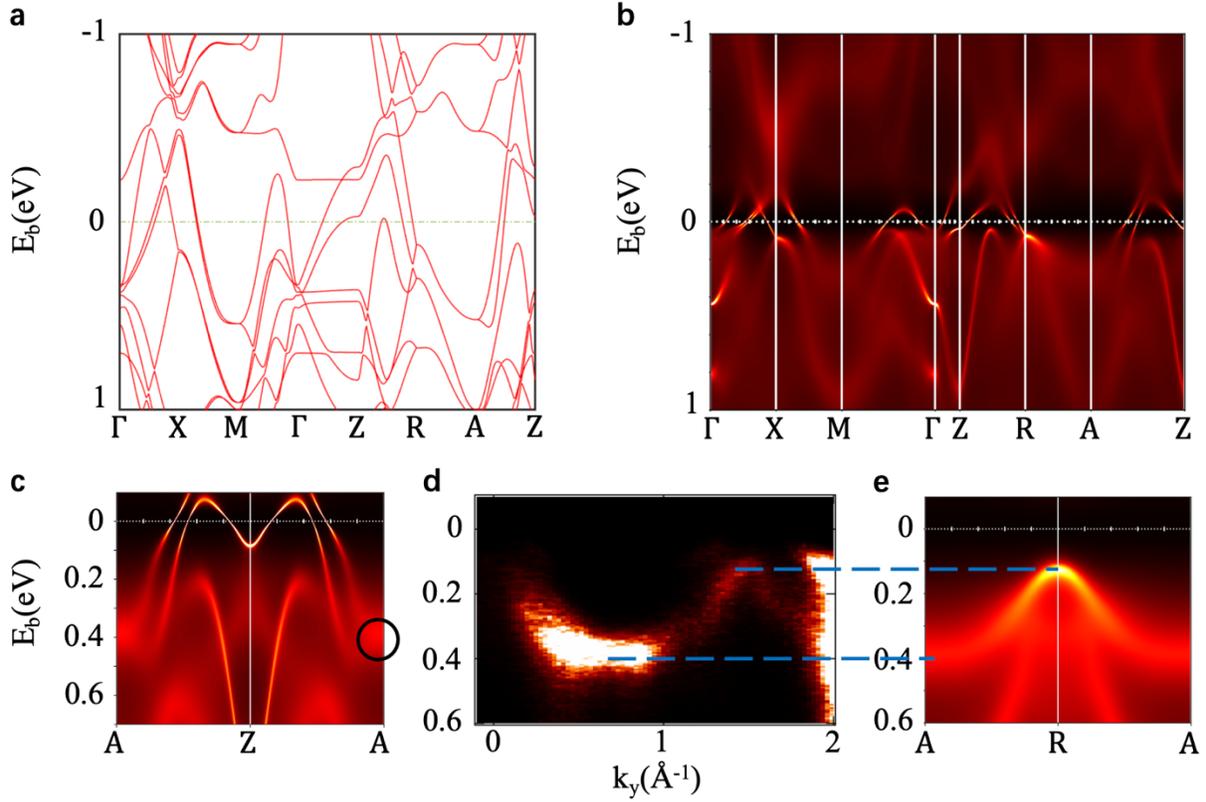

**Fig. 3. YMn$_2$Ge$_2$ as a topological Hund metal. a,** DFT calculation of the bulk electronic structure in YMn$_2$Ge$_2$ with spin-orbit coupling and magnetism included. **b,** DFT+DMFT calculation of the bulk electronic structure in YMn$_2$Ge$_2$. A clear renormalization of the bands near the Fermi level can be resolved compared with **a**. **c,** DFT+DMFT electronic band structure along A-Z-A direction. The black circle indicates the location of the nodal point. **d,** Experimental AFM nodal line along A-R direction (same as Fig. 1**i**). **e,** DFT+DMFT band structure showing an excellent agreement with **d**. The two blue dashed lines confirm the consistency of DFT+DMFT and ARPES data, especially at the R and A points.

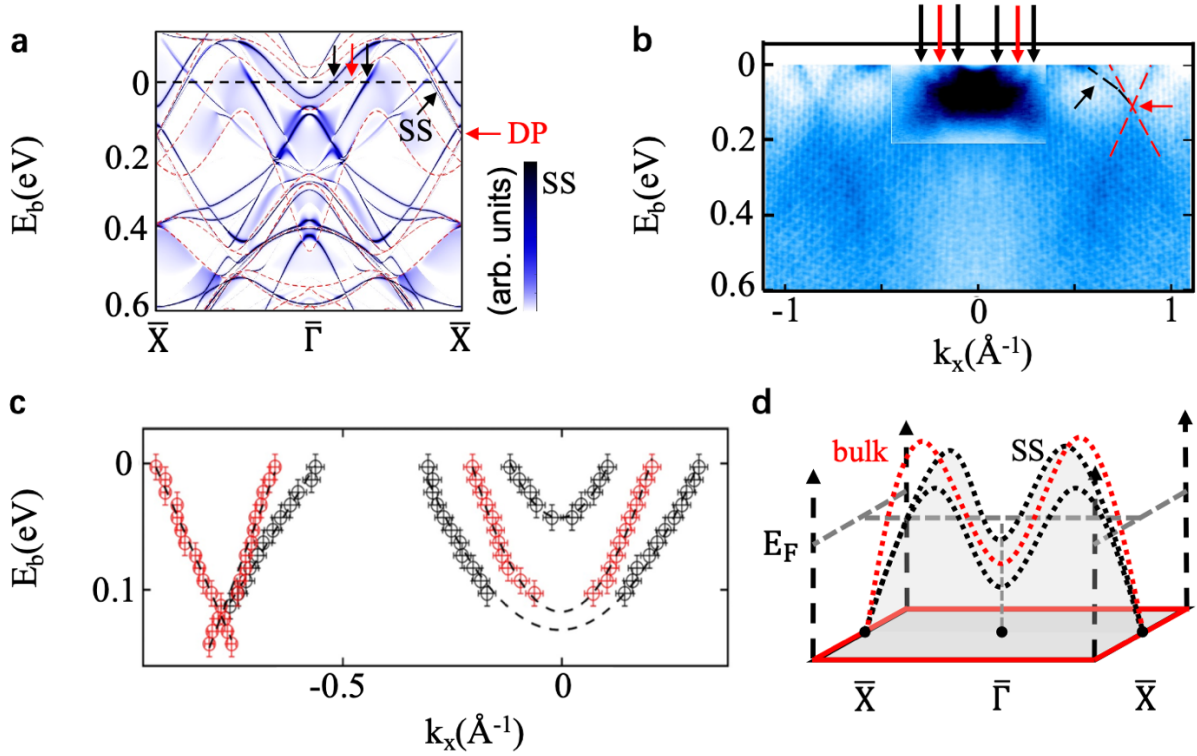

**Fig. 4. Drumhead surface state associated with the AFM nodal line. a,** Semi-infinite surface calculations along the $\bar{X}$-$\bar{\Gamma}$-$\bar{X}$ high symmetry direction. Color bar shows the surface state contribution. Red dashed lines are the bulk states on the A-R-Z plane. Black and red arrows represent the drumhead surface states associated with the AFM nodal line and the bulk state on the A-R-Z plane, respectively. Black dashed line marks the experimental Fermi level. SS and DP mean surface state and Dirac point, respectively. **b,** ARPES dispersion map along the R-Z-R high symmetry direction. A high-resolution dispersion map (same as cut 2 in Fig. 1) at the center of the Brillouin zone (Z point) is embedded to highlight the surface and bulk states. Red and black dashed lines demonstrate the bulk Dirac crossing and the drumhead surface state at R point, respectively. Red arrows are the bulk states in **a** and black arrows represent the drumhead surface state also marked in **a**. **c,** Extracted dispersions of the bulk (red dots) and drumhead surface states (black dots) near the Fermi level. Error bars correspond to the experimental momentum and energy resolutions. Black dashed lines are the linear (left) and quadratic (right) fits to the ARPES data. **d,** Illustration of the energy dispersion of the drumhead surface states near the Fermi level.